\documentclass[12pt]{article}
\begin{document}

\setlength{\unitlength}{1mm}

 \title{Does phantom energy produce black hole?}
 \author{\Large $F.Rahaman^*$ ,  $A. Ghosh^*$ and $M. Kalam^{**}$ }
\date{}
 \maketitle
 \begin{abstract}
                  We have found an exact solution of spherically
                  symmetrical Einstein equations describing a
                  black hole with a special type phantom energy source. It is
                  surprising to note that our solution is
                  analogous to Reissner-Nordstr\"{o}m black hole.

  \end{abstract}



 \bigskip
 \medskip
  \footnotetext{ Pacs Nos : 98.80 cq, 04.20 Gz, 04.40 Nr   \\

                              $*$Dept.of Mathematics,Jadavpur University,Kolkata-700 032,India\\
                                  E-Mail:farook\_rahaman@yahoo.com

                             $**$Dept. of Phys. , Netaji Nagar College for Women ,
                                          Regent Estate,
                                          Kolkata-700092,India
                              }

    \mbox{} \hspace{.2in}
Recent astrophysical observations have confirmed that the
Universe at present is expanding with an acceleration. It is
proposed that this unexpected cosmological behavior is caused by
a hypothetical dark energy with a positive energy density and a
negative pressure. The matter with the property,  energy density
$\rho > 0 $ but pressure $p < 0$ has been denoted 'phantom
energy'. Several authors have recently discussed accelerating
phase of the Universe by using phantom energy as source [1-8].
Since traversable Wormholes require so called exotic matter with a
negative pressure $p < 0$ , some  authors  have recently
investigated the physical properties and characteristic of
traversable Wormholes by taking phantom energy  as source [ 9-13].
In this article, we present a black hole solution with special
type phantom energy as a source. By choosing the parameters
adequately , the solution coincides with Reissner-Nordstr\"{o}m
black hole solution. Before discussing this surprising result, we
proceed to show the solution.

 We look for static spherically
symmetric solution with the line element

\begin{equation}
                ds^2 = - e^{2f(r)} dt^2 + \frac{1}{[1 - \frac{b(r)}{r}]}dr^2+r^2 d\Omega_2^2
            \label{Eq1}
          \end{equation}

\pagebreak
 Because of spherical symmetry the only non zero
components of stress energy tensor are $ T^0_0 = - \rho(r)$, $
T^1_1 = p(r)$ and $T^2_2 = T^3_3 = p_{tr}(r)$ where $\rho$ is the
energy density , $p$ is the radial pressure and $p_{tr}$ is the
tranverse pressure. Using the Einstein field equations
 $G_{\mu\nu} = 8\pi T_{\mu\nu} $, in orthonormal reference frame (
with $ c = G = 1 $), we obtain the following stress energy
scenario,
\begin{equation}
                \rho(r) =\frac{b^\prime}{8\pi r^2}
                \label{Eq3}
          \end{equation}

\begin{equation}
                p(r) =\frac{1}{8\pi} [ - \frac{b}{r^3} + 2\frac{f^\prime}{r} ( 1 -
                \frac{b}{r})]
                \label{Eq3}
          \end{equation}

\begin{equation}
                p_{tr}(r) =\frac{1}{8\pi}( 1 -
                \frac{b}{r}) [ f^{\prime\prime} - \frac{(b^\prime r - b )}{2r(r-b)}f^\prime
              + {f^\prime}^2 + \frac{f^\prime}{ r } - \frac{(b^\prime r -
              b)
              }{2r^2(r-b)}]
                \label{Eq3}
          \end{equation}

Using the conservation of stress energy tensor $ T^{\mu\nu}_{;\nu}
= 0 $, we can obtain the following equation
\begin{equation}
                p^\prime  + f^\prime \rho + ( f^\prime +  \frac{2}{r})p -
                \frac{2}{r}p_{tr} = 0
                \label{Eq3}
          \end{equation}

 From now on , we assume that our source is characterized by the
 special type phantom energy with equation of state that contains a radial
 pressure

 \begin{equation}
               p =  - \rho
                \label{Eq3}
          \end{equation}
we suppose also that pressures are isotropic and
\begin{equation}
               p_{tr} =   \rho
                \label{Eq3}
          \end{equation}

Since only two equations of the system (2) - (4) are independent
, it is convenient to represent them as follows:

 \begin{equation}
               {b^\prime}= {8\pi r^2}\rho(r)
                \label{Eq3}
          \end{equation}

\begin{equation}
                {f^\prime}=\frac{(8\pi p r^3 + b )}{2r(r-b)}
                \label{Eq3}
          \end{equation}
One can find the solution of $\rho$ from (5) by using (6) and (7)
as
\begin{equation}
                \rho(r) =\frac{\rho_0}{r^4}
                \label{Eq3}
          \end{equation}

where $ \rho_{0} $ is an integration constant.

Plugging (10) in (8) and (9), one can get the following solutions
of b and f as

\begin{equation}
                b = A - \frac{8\pi\rho_0}{r}
                \label{Eq3}
          \end{equation}

\begin{equation}
                2f = \ln f_0 [  \frac { r^2 - Ar +
          8\pi\rho_0}{r^2}]
                \label{Eq3}
          \end{equation}
where $ f_{0} $ and A are integration constants.

Rescaling the time coordinate appropriately, the line element
becomes,

\begin{equation}
                ds^2 = - [ 1 -  \frac{A}{r} + \frac {
          8\pi\rho_0}{r^2} ]dt^2 + \frac{1}{[ 1 -  \frac{A}{r} + \frac {
          8\pi\rho_0}{r^2}]} dr^2+r^2 d\Omega_2^2
            \label{Eq1}
          \end{equation}

The structure of this solution is similar to the
Reissner-Nordstr\"{o}m black hole solution.

 In the absence of the source i.e. when  $\rho_0$ is zero ,
then the metric (13) becomes Schwarzschild metric and comparing
with Schwarzschild metric , the constant A can be chosen to be 2M
, M is the mass of the black hole.

\textbf{ Properties of the solution:}

For $ A > \sqrt{32 \pi\rho_0 }$, there are two zeros of $1 -
\frac{A}{r} + \frac {
          8\pi\rho_0}{r^2}$ at $ r = r_\pm$ where

$  r_\pm = \frac {A\pm \sqrt{A^2 -  32\pi\rho_0}} { 2} $ which
correspond to two horizons.

The Kretschmann scalar

$K = R_{abcd}^{abcd} = \frac {4}{r^6} [ ( A -
\frac{32\pi\rho_0}{r})^2 + ( A - \frac{16\pi\rho_0}{r})^2  + ( A
- \frac{8\pi\rho_0}{r})^2] $

is finite at $  r_\pm $ and is divergent at $ r = 0 $, indicating
that $  r_+ $ and $  r_- $ are regular horizons and the
singularity locates at $  r = 0 $.

For $ A = \sqrt{32 \pi\rho_0 }$, the black hole has only one event
horizon located at $  r = \frac{A}{2} $. Thus two horizons $  r_+
$ and $  r_- $ match to form a regular event horizon while $  r =
0 $ is still a singularity.

We also see that if  $ A < \sqrt{32 \pi\rho_0 }$, the solution
does not describe a black hole at all, but, rather a naked
singularity.

\pagebreak Now we find entropy S and Hawking temperature $T_H$ of
the black hole following Hawking's remarkable discovery - the
laws of black hole thermodynamics [14].

 $S =\frac{1}{4}( area ) = \frac{\pi}{4}[A + \sqrt{(A^2 -
 32\pi\rho_0)}]^2$

$ T_H = \frac{1}{4\pi \sqrt {-g_{tt}g_{rr}}}\frac{d}{dr} ( -
g_{tt})|_{horizon} = \frac{1}{\pi}[\frac{\sqrt{(A^2 -
 32\pi\rho_0)}} {[A + \sqrt{(A^2 -
 32\pi\rho_0)}]^2}] $

In the limiting case i.e. when $ A = \sqrt{32 \pi\rho_0 }$, the
black hole exhibits a non zero entropy $ S_0 = \frac{ \pi A }{4}$,
at zero temperature. One may consider it as a result of a dual
symmetric that generates degenerate ground states of black hole.

In conclusion, we give a black hole solution by taking special
type phantom energy as source. The structure and thermodynamic
properties of this black hole is similar to
Reissner-Nordstr\"{o}m black hole. Three possible questions could
there be to our model.

[1] Is any spherically charged distribution of matter has the
same notion of special type phantom energy and obeys equation of
state : $ p =  - \rho ,  p_{tr} =   \rho$ ?

[2] It is argued that apart from the null energy condition
violation, phantom energy possesses a strange property namely,
phantom energy mediates a long range repulsive force [8]. So, how
it is possible to a source ( distribution of matter ) which
produces repulsive force to form a black hole ?

[3] Comparing the metric (13) with Reissner-Nordstr\"{o}m black
hole metric, one can find $ 8 \pi\rho_0 = e^2 $ i.e. $\rho = \frac
{e^2}{8 \pi r^4} $ [ where e is the charge of the matter and
$\rho $ is the matter density of the phantom energy source ]. So,
is it possible to relate gravity with electromagnetic field?

The answer of these questions is under current consideration and
we hope to report this elsewhere.

\pagebreak

        { \bf Acknowledgements }

          F.R is thankful to Jadavpur University and DST , Government of India for providing
          financial support under Potential Excellence and Young
          Scientist scheme .  \\



\begin{thebibliography}{0}
\bigskip
     \bibitem{kg1} R. Cai and A Wang arXiv: hep-th / 0411025


    \bibitem{kg2}  M. Carmeli arXiv: astro-ph / 0111259
    \bibitem{kg3}  M. Turner arXiv: astro-ph / 0108103
    \bibitem{kg4}  R. Caldwell, M. Kamionkoski and N. Weinberg arXiv: astro-ph / 0302506
.
    \bibitem{kg5}  A. Melchiorai , L. Mersini , C. Odman and M.
    Trodden arXiv: astro-ph / 0211522

    \bibitem{kg6}  S. Caroll, M. Hoffman and M. Todden arXiv: astro-ph / 0301273
    \bibitem{kg7} R. Caldwell arXiv: astro-ph / 9908168
    \bibitem{kg7} E. Majerotto , D. Sapone and L. Amendola  arXiv: astro-ph
    /0410543
       \bibitem{kg7} F. Lobo  arXiv: gr-qc / 0502099
            \bibitem{kg7}S. Sushkov  arXiv: gr-qc / 0502084
             \bibitem{kg7} F. Lobo  arXiv: gr-qc / 0506001
              \bibitem{kg7} O. Zaslavskii  arXiv: gr-qc /
              0508057
              \bibitem{kg7} F Rahaman, M Kalam, M Sarker and K Gayen, Phys.Lett.B 633, 161(2006)
 ( e-Print Archive: gr-qc/0512075 )

               \bibitem{kg7} G. Cheng, W. Lin and R. Hsu J. Math.
               Phys.35, 1839 ( 1994)
    \end{thebibliography}
\end{document}